\documentclass[conference]{IEEEtran}

\usepackage{graphicx}
\usepackage{subcaption}
\usepackage{balance}
\usepackage{comment}
\usepackage{amsmath}
\usepackage[bookmarks=false]{hyperref}

\usepackage{balance}
\usepackage{stfloats}

\usepackage{multirow}
\usepackage[table]{xcolor}

\IEEEoverridecommandlockouts
\begin{document}

\title{Energy and Performance Analysis of STTRAM Caches for Mobile Applications}

	\author{\IEEEauthorblockN{Kyle Kuan and
	Tosiron Adegbija}
	\IEEEauthorblockA{Department of Electrical \& Computer Engineering \\ University of Arizona, Tucson, AZ, USA \\
	Email: \{ckkuan, tosiron\}@email.arizona.edu}
	\vspace{-15pt}
	\thanks{This work was supported in part by NSF Grant CAREER 1844952.}}

	
\maketitle
	
\begin{abstract}
Spin-Transfer Torque RAMs (STTRAMs) have been shown to offer much promise for implementing emerging cache architectures. This paper studies the viability of STTRAM caches for mobile workloads from the perspective of energy and latency. Specifically, we explore the benefits of reduced retention STTRAM caches for mobile applications. We analyze the characteristics of mobile applications' cache blocks and how those characteristics dictate the appropriate retention time for mobile device caches. We show that due to their inherently interactive nature, mobile applications' execution characteristics---and hence, STTRAM cache design requirements---differ from other kinds of applications. We also explore various STTRAM cache designs in both single and multicore systems, and at different cache levels, that can efficiently satisfy mobile applications' execution requirements, in order to maximize energy savings without introducing substantial latency overhead.

\end{abstract}



\section{Introduction}
The past few years have witnessed a mobile tipping point wherein more mobile devices (e.g., smartphones and tablets) are being sold and used than all other kinds of computers combined. In addition, mobile devices continue to run increasingly complex applications in line with users' increasing demands for high-performance, low-energy systems. As a result, the processors featured in mobile devices are becoming more sophisticated, with advanced microarchitecture optimizations, such as out-of-order execution, deep pipelines, multi-level cache hierarchies, asymmetric configurations, etc. \cite{cortexA76}. 

While mobile devices contain several components that impact energy and performance, such as the display and radios, the cache subsystem remains one of the most important components of mobile device processors. The cache bridges the processor-memory performance gap, and is consequential to the processor's overall energy efficiency \cite{mittal14}. As such, there is much ongoing research into technologies for enabling energy efficient caches for resource-constrained processors.

An increasingly popular approach for improving caches' energy efficiency involves replacing the traditional SRAM with emerging non-volatile memory (NVM) technologies. The spin-transfer torque RAM (STTRAM) \cite{apalkov13}, especially, is attractive for implementing on-chip memories, due to several characteristics, such as high density, higher reliability (compared to other NVMs), etc. There is also much ongoing research to mitigate STTRAM's overheads of high write latency, high write energy, reliability issues, etc \cite{kang15}. However, to maximize the benefits of STTRAM caches in mobile devices, we must first understand the execution characteristics of mobile applications within the context of STTRAM caches' unique characteristics.

This paper aims to explore and analyze the energy and performance benefits of STTRAM caches for mobile applications. Specifically, we approach our analysis from the perspective of \textit{reduced retention STTRAM caches} \cite{smullen11}. Prior work showed that STTRAM's long write latency and high write energy can be attributed to the long retention time---the duration for which data is maintained in the memory in the absence of power. For caches, the intrinsic STTRAM retention time of up to 10 years is unnecessary, since most cache blocks need to be retained in the cache for no longer than 1$s$ \cite{Jog12}. As such, the retention time can be substantially reduced to mitigate the write overheads. We note that other techniques for addressing STTRAM's write overheads have been proposed,  but we limit the studies herein to reduced retention STTRAM caches, hereafter simply referred to as 'STTRAM caches.'

In order to maximize the benefits of STTRAM caches for mobile applications, the retention time must suffice for the applications' \textit{cache block lifetimes}. Similar to prior studies on SRAM caches \cite{kaxiras01}, we define cache block lifetimes as the duration for which cache blocks must remain in the cache before they are evicted or invalidated. Given that most mobile applications are intrinsically interactive, and thus exhibit execution characteristics that may differ from other non-interactive applications, it is imperative to study STTRAM caches in the context of mobile applications' characteristics. 

In this paper, we analyze mobile applications with respect to their cache block lifetimes, and other execution characteristics (e.g., read-write behavior) that impact how well STTRAM caches are matched to mobile applications' execution needs. Furthermore, we explore the behavior of mobile applications on STTRAM caches in single core and multicore processors, and derive new insights, based on our analysis, on the tradeoffs of STTRAM caches within these processor contexts. 

\begin{figure}[b]
		\vspace{-20pt}
		\centering
		\includegraphics[width=.5\linewidth]{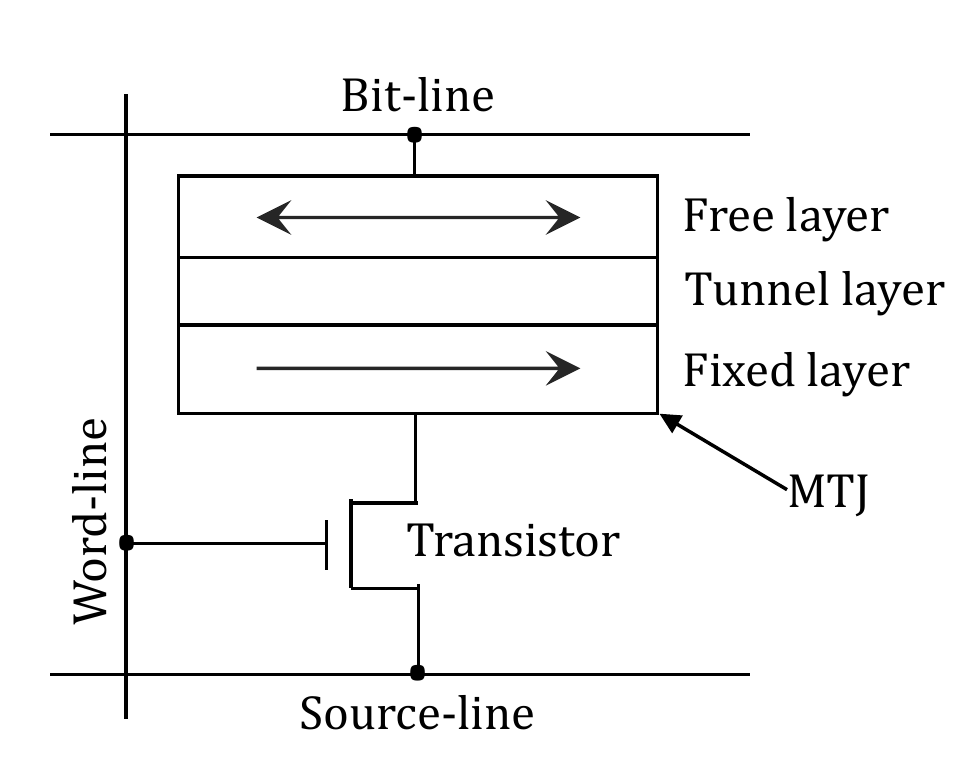}
		\vspace{-7pt}
		\caption{STT-RAM basic cell structure}
		\vspace{-7pt}
		\label{fig:sttram}
	\end{figure}

\section{Background and Related Work}
\subsection{Overview of STTRAM}
Similar to other resistive memories, STTRAM uses non-volatile, resistive information storage in a cell. Fig. \ref{fig:sttram} illustrates STTRAM's basic structure, comprising of a magnetic tunnel junction (MTJ) and a transistor. The MTJ cell, which is used as the binary storage cell, contains two ferromagnetic layers (the free and fixed layers) separated by an oxide barrier/tunnel layer. The free layer's direction with respect to the fixed layer (parallel or anti-parallel) generates low resistance and high resistance states of the MTJ cell, to indicate the "0" or "1" bit. The magnetization change in the free layer is controlled by a transistor, which allows current to flow through the MTJ cell and creates a spin torque that switches the magnetization in the free layer. We direct the reader to \cite{chun13} for additional low-level details of STTRAM's basic structure.

\subsection{Overview of STTRAM-based Analysis}
There has been much prior work on the benefits of replacing SRAM with STTRAM. For instance, Noguchi et al. \cite{noguchi14} showed that replacing SRAM with STT-RAM reduced the last level cache (LLC) energy by 60\%, with a 2\% performance degradation. However, STTRAM still has drawbacks that have slowed down the adoption of STTRAM in emerging processor architectures. Notably, STTRAMs' high write energy and latency overheads have attracted much research attention. One research thread involves relaxing STTRAM's thermal stability to reduce the retention time and thus, the write energy and latency \cite{sun11,smullen11}. Alternatively, prior work has explored other techniques such as dynamic block allocation \cite{wang14} for use in hybrid (SRAM+STTRAM) caches, wherein write-active cache blocks are written into SRAM, while read-active blocks are written into STTRAM. Other techniques involve identifying and eliminating redundant writes or utilizing opportunistic replacement policies in order to minimize the write overheads \cite{ahn14,zhou09,reed17,qiu19}. Recently, Kuan et al. \cite{kuan18} studied the characteristics of different SPEC 2006 benchmarks \cite{spec} and showed that the STTRAM cache energy could be further improved by adapting the retention time to different applications' characteristics, without incurring much optimization overhead. 

Yan et al. \cite{yan17} proposed to partition the L2 cache for user and kernel accesses in mobile devices based on in the variability in access patterns between these two kinds of accesses. However, to the best of our knowledge, there is no prior work that has extensively analyzed the benefits of reduced retention STTRAM caches for mobile applications. Most of the aforementioned prior works used desktop or high performance benchmark suites like SPEC2006 and PARSEC \cite{parsec} in their analysis. Given that STTRAMs offer multiple advantages (e.g., low leakage, high density) for resource-constrained systems, we anticipate that STTRAM caches will play an important role in emerging mobile computing systems. Furthermore, since mobile applications' execution characteristics differ drastically from traditional benchmarks, due in part to mobile applications' interactive nature \cite{gutierrez11}, it is imperative to analyze STTRAM caches with relevant benchmarks that represent mobile applications' characteristics \cite{yan17}.

\section{Methodology}
This section describes our methodology for gathering the data for the analysis presented herein. We first describe some design assumptions made, and thereafter, present our experimental setup.

\subsection{Design Assumptions}
We assume that STTRAM caches can be fabricated as desired with different retention times. We note that reducing the retention time trades off other factors like the reliability \cite{WKang15}. As such, there is much on-going research to address this and other fabrication and device challenges of STTRAMs (e.g., process variation) \cite{senni17}, but addressing these challenges is outside the scope of this paper. This subsection summarizes modeling techniques that we have employed for achieving reduced retention times and for preventing data corruption in reduced retention STTRAM caches. 

\subsubsection{Achieving Reduced Retention Times}
Prior work showed that STTRAM's thermal stability ($\Delta$) can be substantially reduced to lower the write energy and latency \cite{smullen11}. This is more so beneficial for storage media that don't require long retention, such as CPU caches. Thus, to model the STTRAM caches analyzed in this paper, we followed the technique proposed in \cite{chun13}. We decreased the MTJ's planar area to obtain the desired retention times and lower thermal stability, and used the models described in \cite{chun13} to determine the MTJ characteristics for different retention times. Based on these characteristics, we calculated write pulse, write current, and MTJ resistance values $R_{AP}$ and $R_P$. 

\subsubsection{Preventing Data Corruption}
A challenge that arises after reducing STTRAM cache's retention time is that some cache blocks may need to remain in the cache beyond the cache's predetermined retention time. As such, the data could become unstable or corrupt, and could cause incorrect results if reused by the CPU. Thus, to prevent data corruption, we incorporate a per-block counter to keep track of the cache blocks' lifetimes \cite{Jog12,kaxiras01}. The counter detects the expiration of a cache block and evicts the block just before the retention time elapses. Dirty blocks are first written to main memory before eviction. We implemented the counter as a finite state machine, with a clock period defined as the retention time divided by $N$, where $N$ dictates the granularity of block eviction. When a block is written to the cache, the counter's state advances from the initial state until it reaches the maximum state. The block is then evicted, and the counter is reset to the initial state whenever a new fetch operation occurs for the block. We note that this is a low overhead technique that only requires a few bits per block. The implementation in our experiments only required two bits per block for a four-state ($N$ = 4) counter.

\subsection{Experimental Setup and Workloads}

For the analysis presented herein, we used an in-house modified version of the GEM5 simulator \cite{gem5}. The modified GEM5\footnote{The modified GEM5 version can be found at \url{www.ece.arizona.edu/tosiron/downloads.php}} models the behavior of relaxed retention STTRAM caches with specified retention times. We modeled single- and quad-core processors with base configurations similar to those featured in modern mobile devices (e.g., ARM Cortex-A76). The processor featured a 1.9GHz clock frequency, out-of-order execution, private level one (L1) instruction and data caches, and shared unified L2 caches (for the multicore experiments). The L1 caches had 32KB size, 4-way set associative, and 64B line size, while the L2 cache had 2MB size, 16-way set associative, and 64B line size. 

We considered retention times ranging from 1$\mu s$ to 100$ms$ in *10 increments. Later in our analysis, we focus on only the 1$ms$, 10$ms$, and 100$ms$ retention times, since we found that smaller retention times were severely under-provisioned for all the mobile applications considered with respect to energy and latency. To model STTRAM and SRAM cache energy, we used NVSim \cite{dong12} integrated with the GEM5 statistics. We used the MTJ cell modeling technique proposed in \cite{chun13} to obtain essential parameters, such as the write pulse, write current, and MTJ resistance value $R_{AP}$, and then applied these parameters to NVSim to construct the STTRAM caches. 

To represent mobile applications, we used ten benchmarks from the \textit{Moby} benchmark suite \cite{huang14}. Moby comprises of a variety of popular android applications, including web browser, social networking, email client, music player, map, document processing, etc., all of which are selected from the Google Play Store. The benchmarks were run in GEM5 Full System mode using Android Ice Cream Sandwich (ICS) as the operating system. We ran all the benchmarks to completion after skipping the boot process using checkpointing. 


\section{STTRAM-Aware Characterization}

In the analysis presented in this section, we focus on STTRAM-specific insights, since the mobile applications have been analyzed with respect to SRAM caches in prior work \cite{huang14}. We first analyze the characteristics of mobile applications in the context of STTRAM caches, with respect to the \textit{read-write activity}, \textit{cache block lifetimes}, and \textit{expiration misses}; these characteristics directly affect the applications' performance on STTRAM caches. In the following section, we analyze the energy and performance (latency) of mobile applications in different STTRAM cache design scenarios.

\subsection{Read-Write Activity}

\begin{figure}[t]
		\vspace{0pt}
		\centering
		\includegraphics[width=0.7\linewidth]{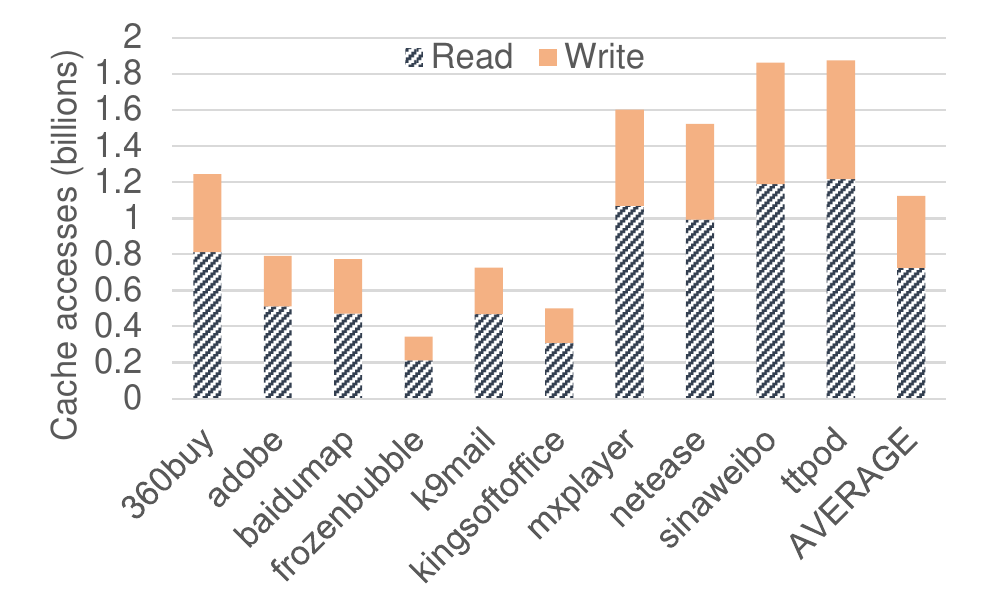}
		\vspace{-7pt}
		\caption{Read-write ratio for data cache memory accesses}
		\label{fig:read-write}
        \vspace{-10pt}
\end{figure}

An application's read-write activity---the ratio of reads to writes---can substantially affect the efficiency of executing those applications on STTRAM memories. Since STTRAMs suffer the most overheads during write operations (due to the high write energy and latency), applications with more reads than writes tend to benefit more from STTRAM caches. Thus, we analyzed the read-write behavior of the mobile applications in the context of single and multicore systems.

Figure \ref{fig:read-write} depicts the read-write ratio of data cache accesses (in billions of instructions) of the mobile benchmarks in a single core execution. Even though the total number of accesses differed substantially across the different applications, we observed that the read-write ratio was relatively stable across the applications. Overall, the reads were, on average, 67\% of total accesses, ranging from 66\% to 70\%. 

It is possible for an application's threads to exhibit different read-write behaviors when they are distributed across different cores vs when the application is running on a single core. Thus, we also analyzed the read-write behaviors across different threads in a multicore scenario. The trends were consistent across all the threads running on different cores: on average across all the benchmarks, the reads were 64\% of the total accesses, ranging from 59\% to 73\%. This high proportion of reads suggests that the considered mobile applications, in general, would consistently be less susceptible to the overheads of STTRAM caches than applications that exhibit high variability in the read-write ratios, such as SPEC benchmarks \cite{Jog12}.

\begin{figure}[t]
		\vspace{0pt}
		\centering
		\includegraphics[width=0.6\linewidth]{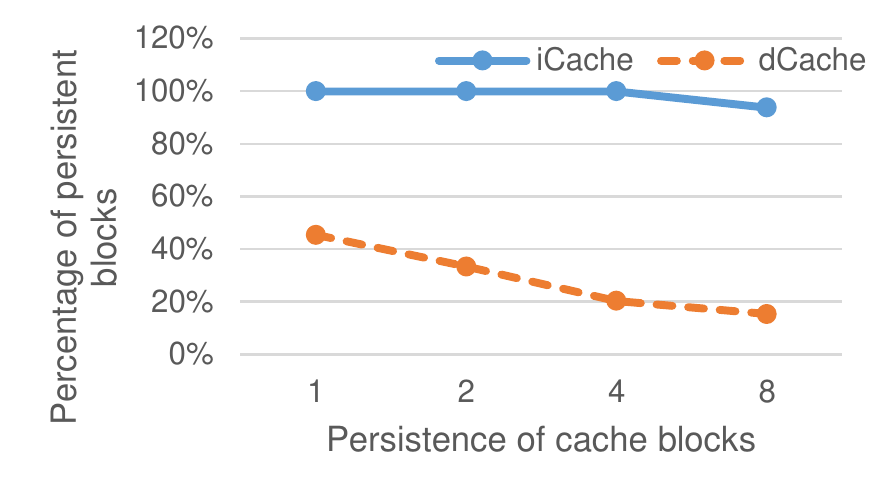}
		\vspace{-7pt}
		\caption{Percentage of persistent blocks. The results depicted are on average across all the benchmarks.}
		\label{fig:persistence}
        \vspace{-10pt}
\end{figure}

\subsection{Cache Block Lifetimes and Persistence} \label{sec:cacheBlockLifetimes}

Cache block lifetimes and reuse are interrelated and refer to how long a cache block must remain in the cache before it is evicted or invalidated. The cache block lifetime directly affects an application's retention time requirements and, in effect, an STTRAM cache's performance for the application. We studied the mobile applications' cache block lifetimes to derive insights into their retention time requirements. We observed that, on average, the cache block lifetimes of the mobile applications varied across the different applications, but mostly ranged from 1$ms$ to 100$ms$. These observations about the cache block lifetimes dictate the retention times that perform best for the different applications, as will become more evident in subsequent sections.

To further understand the behavior of mobile applications in the context of STTRAM caches, we analyzed the percentage of unique cache blocks that were persistent, given a \textit{persistence threshold (thd)}. We define a block as persistent if it is loaded into cache at least \textit{thd} times after it is evicted from the cache. The persistence of a cache block can provide a sense of how long the block must \textit{ideally} remain in the cache to service all its future references. We analyzed the blocks' persistence for both instruction and data caches with \textit{thd} values of 1 to 8, in power-of-two increments.

Figure \ref{fig:persistence} depicts the persistence of instruction and data cache blocks for persistence thresholds of 1, 2, 4, and 8. In the instruction cache, 100\% of cache blocks were loaded more than four times, and 94\% were loaded more than eight times, revealing a high degree of persistence. In the data cache, on the other hand, only 45\% of cache blocks were loaded more than once, and 15\% were loaded more than eight times. We also observe a much sharper drop in the blocks' persistence as the persistence threshold increases from 1 to 8 for the data cache. On average across all the benchmarks, the data cache had substantially more (72x, on average) unique blocks referenced than the instruction cache. However, the instruction cache's total accesses exceeded the data cache's by more than 2x, due to the high persistence of the instruction cache blocks. 

\begin{figure}[t]
		\centering
		\includegraphics[width=\linewidth]{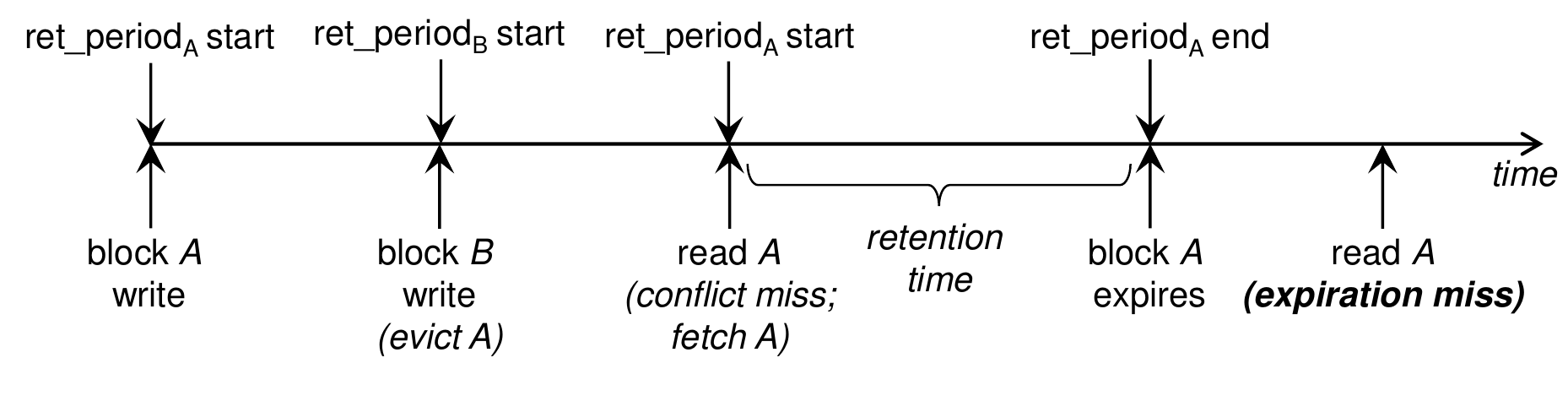}
		\vspace{-10pt}
		\caption{Illustration of expiration miss.}
		\label{fig:expiration}
\end{figure}

\subsection{Expiration Misses} \label{sec:expiration}
In STTRAM caches, an important characteristic that can significantly impact the energy and latency is what we call the \textit{expiration misses.} The expiration misses are introduced by reduced retention times and refer to the misses that result from references to a block that was prematurely evicted due to elapsed retention time. Figure \ref{fig:expiration} illustrates the occurrence of an expiration miss. Assuming a block $A$ and $B$ reside in the same cache set location, and a write of $A$, followed by a write of $B$ evicts $A$, a subsequent read request for $A$ would result in a conflict miss. $A$ is then fetched and its retention period is started. When the retention time elapses, $A$ expires, and a subsequent reference to $A$ results in an expiration miss.

\begin{figure}[t]
		\centering
		\includegraphics[width=0.9\linewidth]{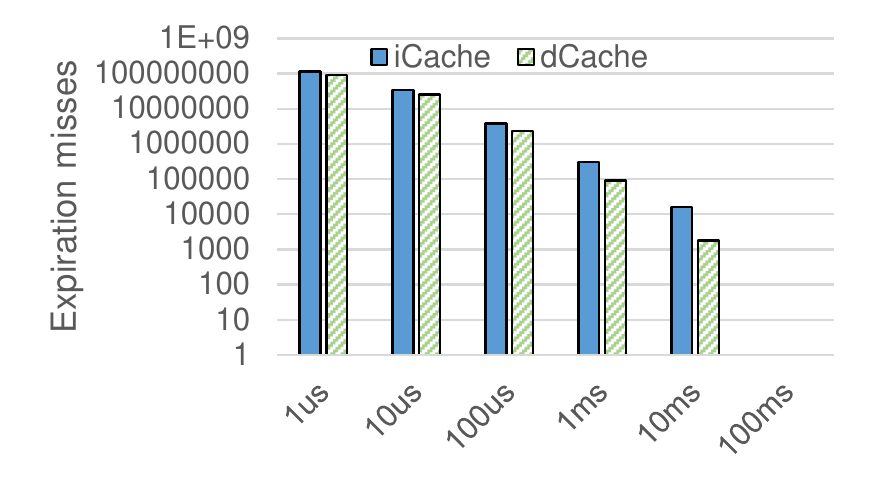}
		\vspace{-7pt}
		\caption{Average expiration misses for different retention times.}
		\label{fig:expMisses}
\end{figure}

Figure \ref{fig:expMisses} depicts the average number of expiration misses for different retention times. The figure depicts the average across all the benchmarks, since the trends were similar for the different benchmarks. Unsurprisingly, as the retention time increased, the number of expiration misses decreased substantially. However, we observed a much sharper decrease for the data cache than for the instruction cache. For instance, the decrease from 1$\mu s$ to 10$\mu s$, 10$\mu s$ to 100$\mu s$, 100$\mu s$ to 1$ms$, and 1$ms$ to 10$ms$ were approximately 3x, 9x, 13x, and 19x, respectively, for the instruction cache; for the data cache, the decreases were 4x, 11x, 25x, and 51x, respectively. Also, there were substantially more expiration misses in the instruction cache than the data cache, and the difference increased as the retention time increased---ranging from 1.2x at 1$\mu s$ to 9x at 10$ms$. For both the instruction and data caches, the number of expiration misses was 0 on the 100$ms$ retention time. 

In general, expiration misses of zero is ideal. However, we also found that the expiration misses is an insufficient criterion for evaluating the benefits of STTRAM cache and different retention times for executing applications. Some applications that have low persistence blocks as majority may tolerate expiration misses better than others, depending on the kinds of misses that occurred. An increase in expiration misses did not necessarily increase energy or latency. For instance, a single miss may accrue less energy and latency overhead than another miss due to future accesses that depend on the miss. Overall, we found that most of the applications were able to tolerate some expiration misses. As such, even though 100$ms$ resulted in zero expiration misses, it was not necessarily the best for all applications, as will become clear in the following sections.

\section{Single Core Evaluation} \label{sec:singlecore}

\begin{figure}[t]
\centering
\begin{subfigure}[t]{.258\textwidth}
  \centering
  \includegraphics[width=\linewidth]{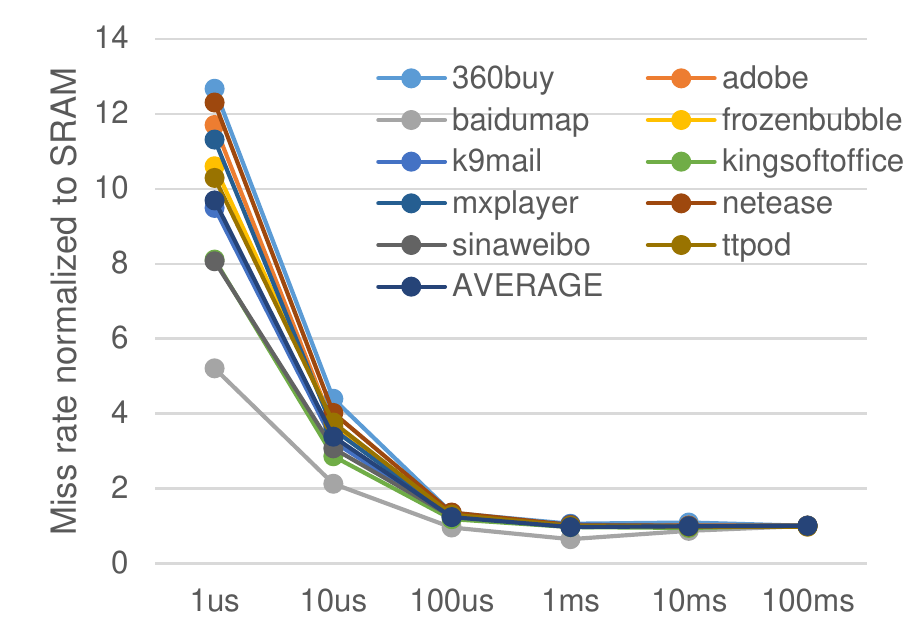}
  \caption{Instruction cache}
  \label{fig:icache_miss-1}
\end{subfigure}%
~
\begin{subfigure}[t]{.258\textwidth}
  \centering
  \includegraphics[width=\linewidth]{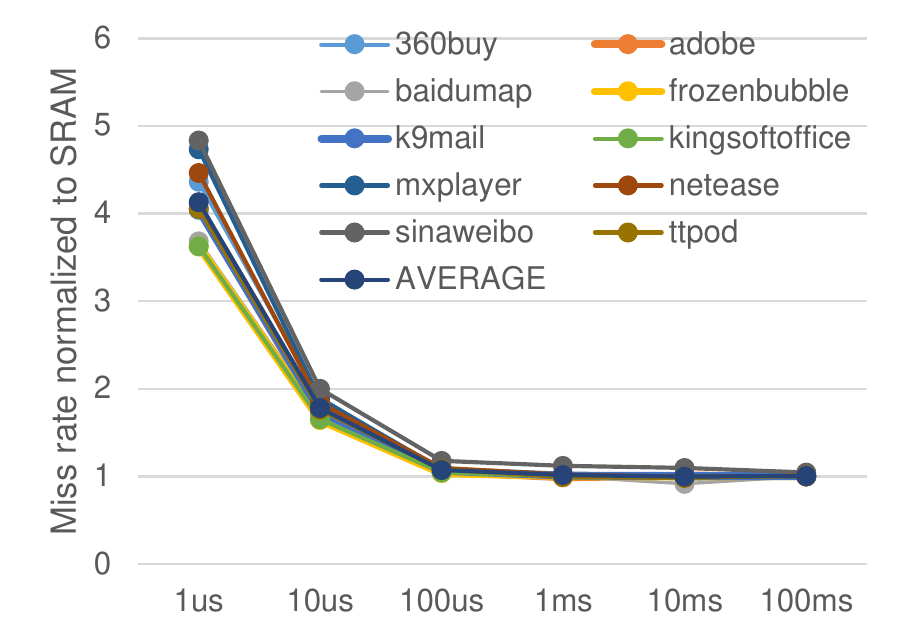}
  \caption{Data cache}
  \vspace{-7pt}
  \label{fig:dcache_miss-1}
\end{subfigure}
\caption{Instruction and data cache miss rates for different retention times normalized to SRAM}
\label{fig:miss-1-core}
\vspace{-15pt}
\end{figure}

\subsection{Cache Miss Rates and Energy}

First, we analyze the benefits of STTRAM caches in the context of a single core processor. Since mobile applications typically execute in energy-constrained environments (i.e., battery-powered mobile devices), the goal is an STTRAM cache that minimizes the energy without substantially increasing the latency compared to SRAM. Thus, we first explore how the instruction and data cache miss rates and energy consumption change for different retention times, and in comparison to SRAM. 

Figure \ref{fig:miss-1-core} depicts the instruction and data cache miss rates of different retention times normalized to SRAM. As expected, as the retention time increased, the cache miss rates trended towards SRAM for both the instruction and data caches. Importantly, we observed that both the data (Figure \ref{fig:dcache_miss-1}) and instruction (Figure \ref{fig:icache_miss-1}) caches exhibit high variability in cache miss rates. That is, different applications require different retention times---for both the instruction and data caches---to achieve the lowest miss rates. This observation is contrary to prior that studied SPEC benchmarks and showed that variability only exists in the data cache, while a single retention time sufficed for the instruction cache across all the SPEC benchmarks \cite{kuan18}. We attribute our observation to the interactive nature of mobile applications, which introduces variability to the instruction cache behavior \cite{gutierrez11}. 

\begin{figure}[t]
\centering
\begin{subfigure}[t]{.258\textwidth}
  \centering
  \includegraphics[width=\linewidth]{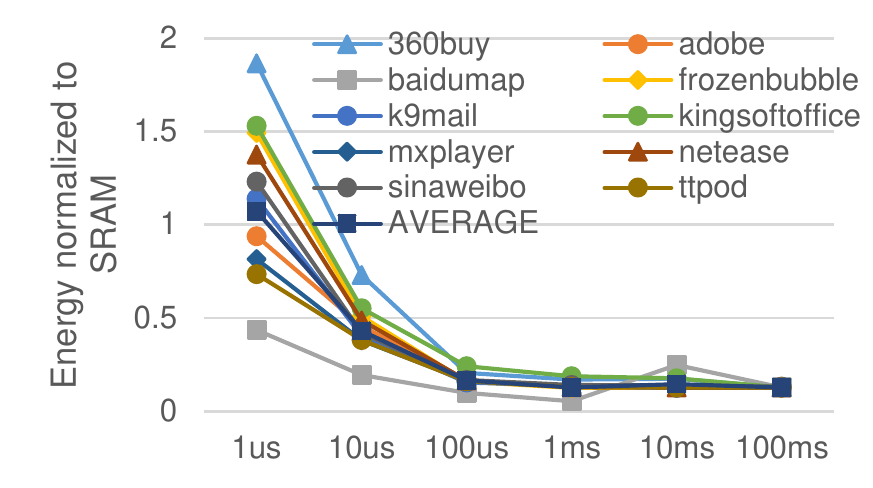}
  \caption{Instruction cache}
  \label{fig:icache-1-E}
\end{subfigure}%
~
\begin{subfigure}[t]{.258\textwidth}
  \centering
  \includegraphics[width=\linewidth]{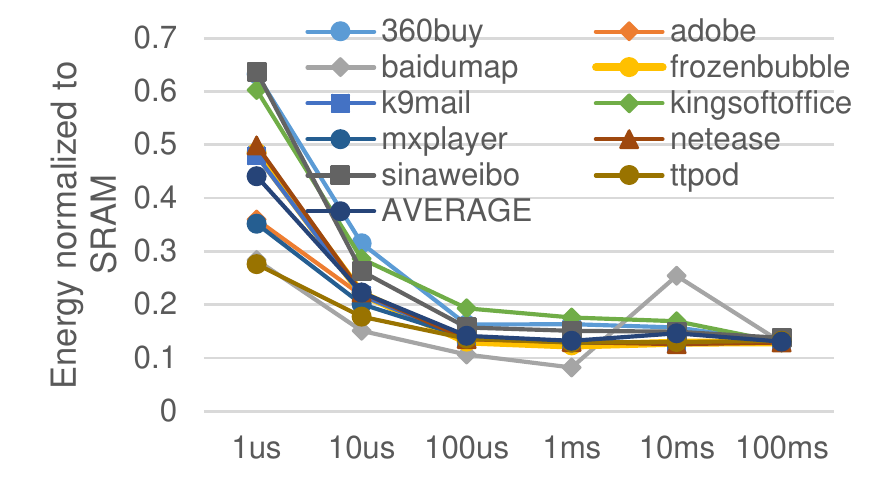}
  \caption{Data cache}
  \vspace{-7pt}
  \label{fig:dcache-1-E}
\end{subfigure}
\caption{Instruction and data cache energy consumption for different retention times normalized to SRAM}
\label{fig:1-core-E}
\end{figure}

For the energy consumption, we observed similar trends between the instruction and data caches. However, there were a few differences. Figure \ref{fig:1-core-E} depicts the instruction and data cache energy consumption trends for different retention times normalized to SRAM. Even though lower retention times would consume less energy per access than SRAM, the behavior is different in the context of application execution. For both the instruction (Figure \ref{fig:icache-1-E}) and data (Figure \ref{fig:dcache-1-E}) caches, the smallest retention times (1$\mu s$ and 10$\mu s$) substantially increased the cache accesses, and in effect the energy consumption. 

In the instruction cache, despite the substantial energy savings of low retention times, the cache accesses due to expiration misses (Section \ref{sec:expiration}) were so substantially increased that the energy consumption at 1$\mu s$ was \textit{more} than SRAM for some applications (e.g., \textit{$360buy, kingsoftoffice, netease$,} etc.). For instance, on average\footnote{We used the geometric mean for the averages discussed in the evaluations.}, the 1$\mu s$ retention time \textit{increased} the energy consumption by 7.1\%, and by up to 86.4\% for $360buy$. The average energy savings for the other retention times ranged from 57\% for 10$\mu s$ to 87.3\% for 1$\mu s$, with little variance for the 100$\mu s$, 1$ms$, 10$ms$, and 100$ms$ retention times. For the data cache, on the other hand, none of the retention times exceeded SRAM in energy consumption despite the increase in cache accesses at lower retention times. The average energy savings ranged from 55.9\% for 1$\mu s$ to 87\% for 100$ms$.

These results illustrate the variable retention time needs of different mobile applications. Due to the cache block lifetimes (Section \ref{sec:cacheBlockLifetimes}), the best retention times for energy ranged from 1$ms$ to 100$ms$, with a majority at 10$ms$---five and seven benchmarks for the instruction and data cache, respectively. None of the benchmarks required less than 1$ms$ for either cache for minimum energy.  

\begin{figure}[t]
\centering
\begin{subfigure}[t]{.258\textwidth}
  \centering
  \includegraphics[width=\linewidth]{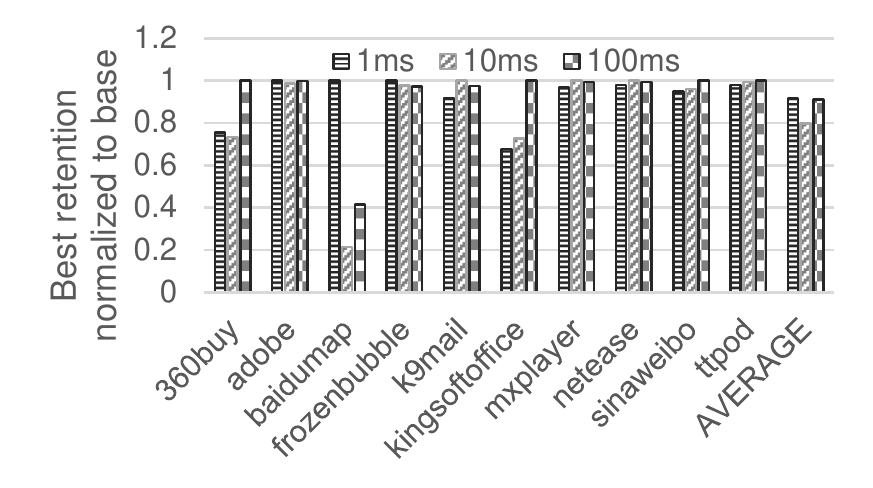}
  \caption{Cache energy}
  \vspace{-7pt}
  \label{fig:adapt-1-iCache-E}
\end{subfigure}%
~
\begin{subfigure}[t]{.258\textwidth}
  \centering
  \includegraphics[width=\linewidth]{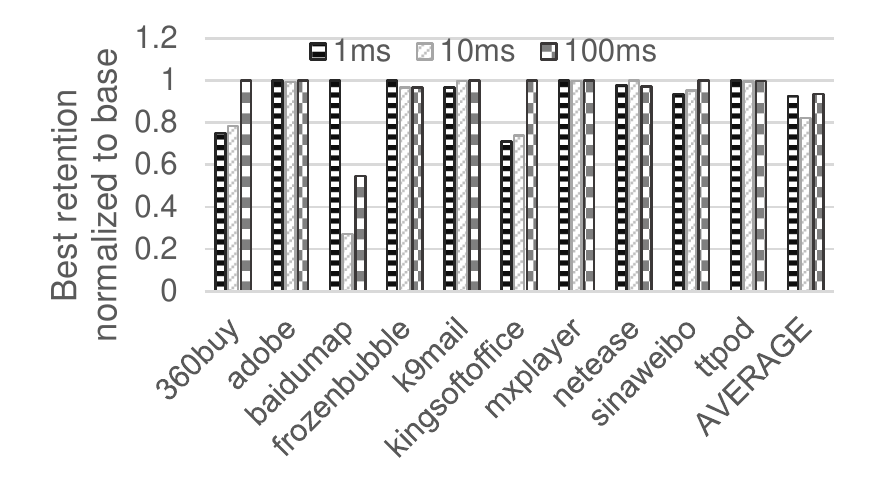}
  \caption{Execution time}
  \label{fig:adapt-1-time}
\end{subfigure}
\caption{Benefits of variable retention times. The best retention time for (a) energy and (b) execution time for each benchmark is normalized to a base of 1$ms$, 10$ms$, and 100$ms$. The instruction and data caches exhibited similar energy trends.}
\label{fig:adapt-1}
\end{figure}

\subsection{Benefits of Retention Time Specialization} \label{sec:retSpecialization}
Prior work (e.g., \cite{kuan18}) suggested using a variety of cache units with different retention times in a single chip to enable close specialization to applications' retention time needs. Thus, we also explored the benefits of specializing the retention time to individual benchmarks' needs vs. using a base configuration of 1$ms$, 10$ms$, or 100$ms$. We selected these three retention times as base, since they covered the best options for all the benchmarks. A design option for such a variable retention time system is described in \cite{kuan18}, wherein the cache chip is designed with different cache units featuring different retention times to satisfy a variety of applications' needs. During runtime, the applications' cache accesses are serviced by the cache unit that best satisfies the applications' needs as determined via sampling or a tuning algorithm. 

Figure \ref{fig:adapt-1} depicts the benefits of such a system in a single core. We assumed a design similar to \cite{kuan18}, with a sampling technique, which samples the application on each cache unit for a sampling interval of 10M instructions to determine the best retention time for different objective functions (energy or latency). The energy savings were similar for both the instruction and data caches. On average across the benchmarks, the best retention times improved the energy for the instruction cache by 8.5\%, 20.2\%, and 9.1\%, compared to the 1$ms$, 10$ms$, and 100$ms$ base retention times, respectively. The data cache energy savings were 7\%, 16\%, and 5.7\%, respectively (Figure \ref{fig:adapt-1-iCache-E} is used to illustrate the cache energy due to the similar trends). As depicted in Figure \ref{fig:adapt-1-time}, on average across the benchmarks, the best retention times improved the performance (execution time) by 7.3\%, 17.7\%, and 6.5\%, respectively. 

\section{Multicore Evaluation} \label{sec:multicore}

\subsection{Single Level Cache}

Next, we explore the benefits of STTRAM caches in a quad-core system with a single level cache. We assumed that a single multi-threaded mobile application was running with the threads distributed among the four cores. Similar to the single core (Figure \ref{fig:miss-1-core}), as the retention time increased, the miss rates trended toward SRAM. Overall, no retention time below 1$ms$ sufficed for any of the considered applications. For instance, for the instruction cache, the 100$\mu s$ retention time increased the cache miss rates by 30.3\%, on average, whereas 1$ms$ achieved similar cache miss rates to SRAM. We observed similar behaviors for the data cache. Thus, we limit the following discussions to only include the 1$ms$, 10$ms$, and 100$ms$ retention times.

\begin{figure}[t]
\centering
\begin{subfigure}[t]{.25\textwidth}
  \centering
  \includegraphics[width=\linewidth]{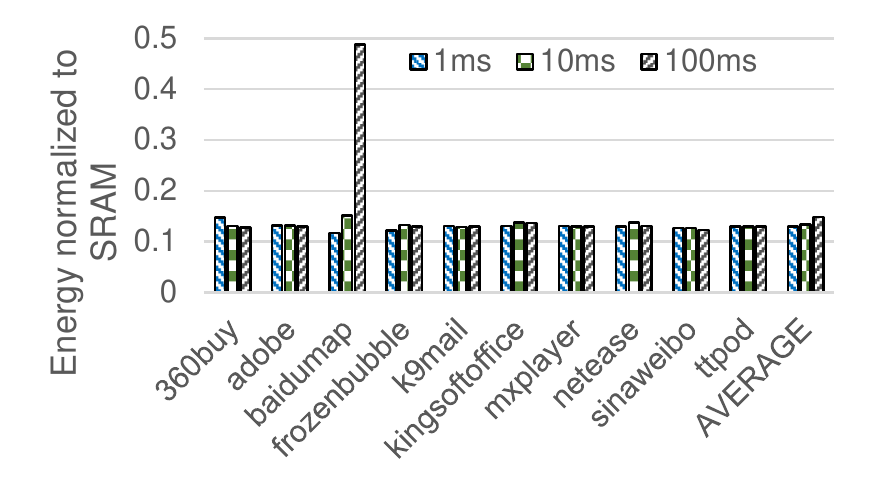}
  \caption{Total energy}
  \label{fig:energy-4}
\end{subfigure}%
~
\begin{subfigure}[t]{.25\textwidth}
  \centering
  \includegraphics[width=\linewidth]{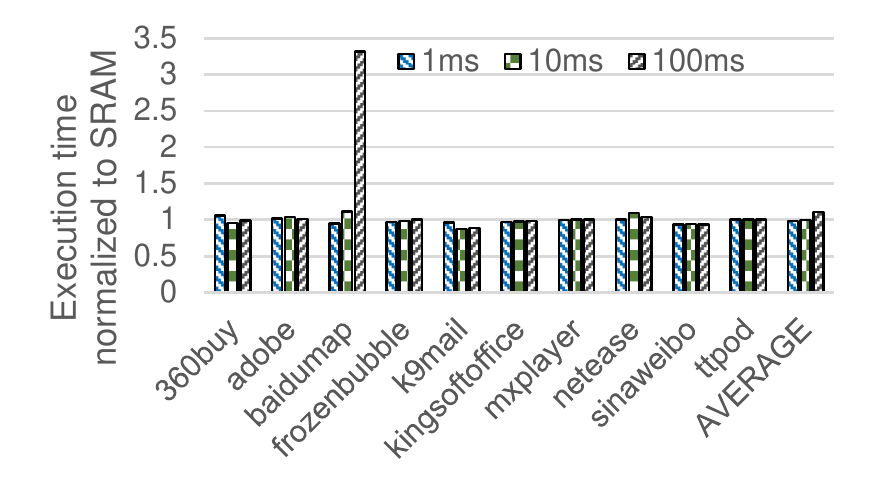}
  \caption{Execution time}
  \vspace{-7pt}
  \label{fig:time-4}
\end{subfigure}
\caption{Energy and execution time of STTRAM cache normalized to SRAM in a quad-core system}
\label{fig:4-core}
\vspace{-15pt}
\end{figure}

\begin{figure*}[hb]
\vspace{-10pt}
\centering
\begin{subfigure}[b]{.30\textwidth}
  \centering
  \includegraphics[width=\linewidth]{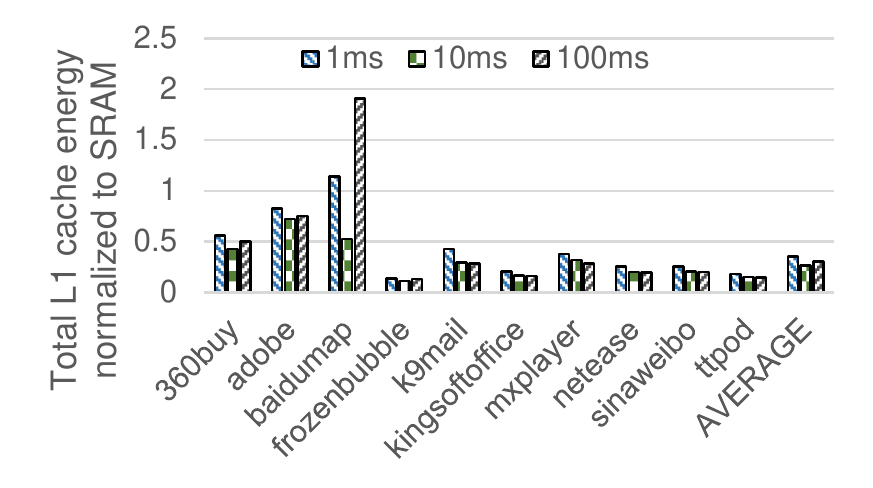}
  \caption{Total L1 cache energy}
  \vspace{-7pt}
  \label{fig:energy-4-L2}
\end{subfigure}%
~
\begin{subfigure}[b]{.3\textwidth}
  \centering
  \includegraphics[width=\linewidth]{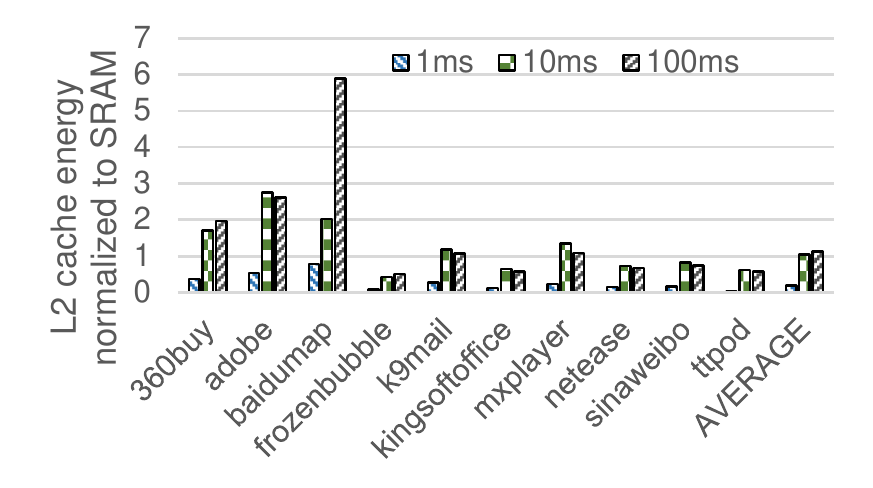}
  \caption{L2 cache energy}
  \vspace{-7pt}
  \label{fig:energy-L2}
\end{subfigure}
~
\begin{subfigure}[b]{.3\textwidth}
  \centering
  \includegraphics[width=\linewidth]{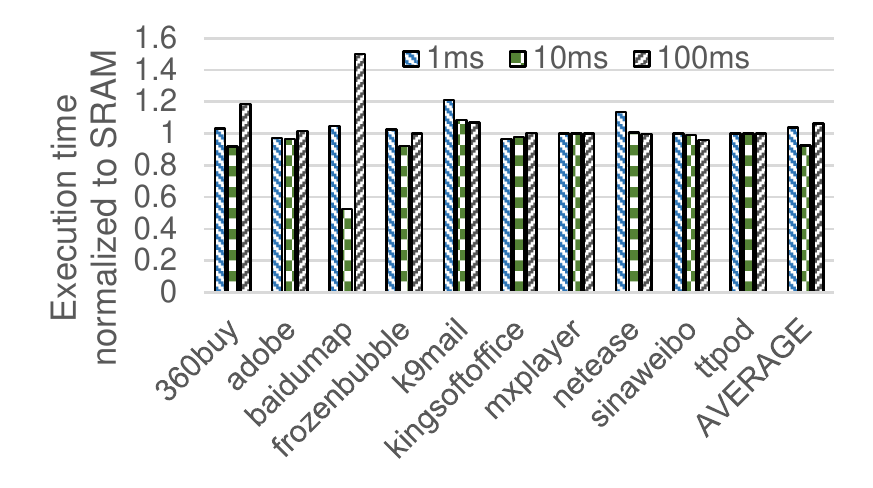}
  \caption{Execution time}
  \vspace{-7pt}
  \label{fig:time-4-L2}
\end{subfigure}
\caption{Total STTRAM L1 and L2 cache energy and time normalized to SRAM. We assume homogeneous STTRAM retention times across all cache levels}
\label{fig:energy-L1-L2}
\vspace{-15pt}
\end{figure*}

Just like the single core system, the multicore STTRAM caches achieved substantial energy savings compared to SRAM. We observed similar trends for both the instruction and data caches, so show results for total cache energy savings.  Figure \ref{fig:energy-4} depicts the total cache energy of STTRAM normalized to SRAM in a quad-core system. On average across all the benchmarks, 1$ms$, 10$ms$, and 100$ms$ STTRAM caches reduced the energy by 87.0\%, 86.6\%, and 85.2\%, respectively, compared to SRAM, with energy savings ranging from 82\% to up to 87\%. The energy savings were consistent across the benchmarks, except for $baidumap$, which was an outlier on the 100$ms$ retention time---due to its sensitivity to the increased write access energy---with 51.3\% energy savings.

The energy savings was at the expense of some execution time overhead. Figure \ref{fig:time-4} depicts the execution time of the quad-core system with STTRAM normalized to SRAM. On average across all the benchmarks, the 1$ms$ and 10$ms$ retention times achieved similar execution times to SRAM. However, the 100$ms$ retention time \textit{increased} the execution time by 10.8\%, owing to \textit{baidumap}, for which the execution time was substantially higher (3x) than SRAM, due its sensitivity to the increased write access latency.

\subsection{Two-Level Cache} 

The results were substantially different when an L2 cache was incorporated into the system. Figure \ref{fig:energy-L1-L2} depicts the total STTRAM L1 energy (\ref{fig:energy-4-L2}), L2 energy (\ref{fig:energy-L2}), and execution time (\ref{fig:time-4-L2}) normalized to SRAM.  When the L2 cache was incorporated, the energy savings in the L1 caches reduced significantly from the system without L2. As shown in Figure \ref{fig:energy-4-L2}, on average, the 1$ms$, 10$ms$, and 100$ms$ L1 STTRAM caches' energy savings were 64.9\%, 73.4\%, and 69.3\%, respectively, compared to SRAM. For the L2 cache, as shown in Figure \ref{fig:energy-L2}, on average, the 1$ms$ STTRAM L2 cache reduced the energy by 80.7\% compared to SRAM. However, the 10$ms$ and 100$ms$ caches \textit{increased} the average energy by 4.4\% and 12.1\%, respectively, compared to SRAM. This was due to the increase in write latency of the 10$ms$ and 100$ms$ caches. 

Figure \ref{fig:time-4-L2} depicts the execution time of the quad-core system featuring L1 and L2 STTRAM caches normalized to SRAM. On average, the 1$ms$ and 100$ms$ STTRAM caches \textit{increased} the execution time by 3.7\% and 6.4\%, respectively, while the 10$ms$ reduced the execution time by 7.6\%. The 10$ms$ cache provided a balance between the overhead of expiration misses in the 1$ms$ cache and the write latency of the 100$ms$ cache. We also note that the STTRAM system with the L2 cache improved the execution time compared to the system without the L2 cache by 8.2\%, 12.4\%, and 12.1\% for the 1$ms$, 10$ms$, and 100$ms$ caches, respectively. These performance improvements were at the expense of increased energy consumption incurred by introducing the L2 cache.

\begin{figure}[t]
		\vspace{-10pt}
		\centering
		\includegraphics[width=0.8\linewidth]{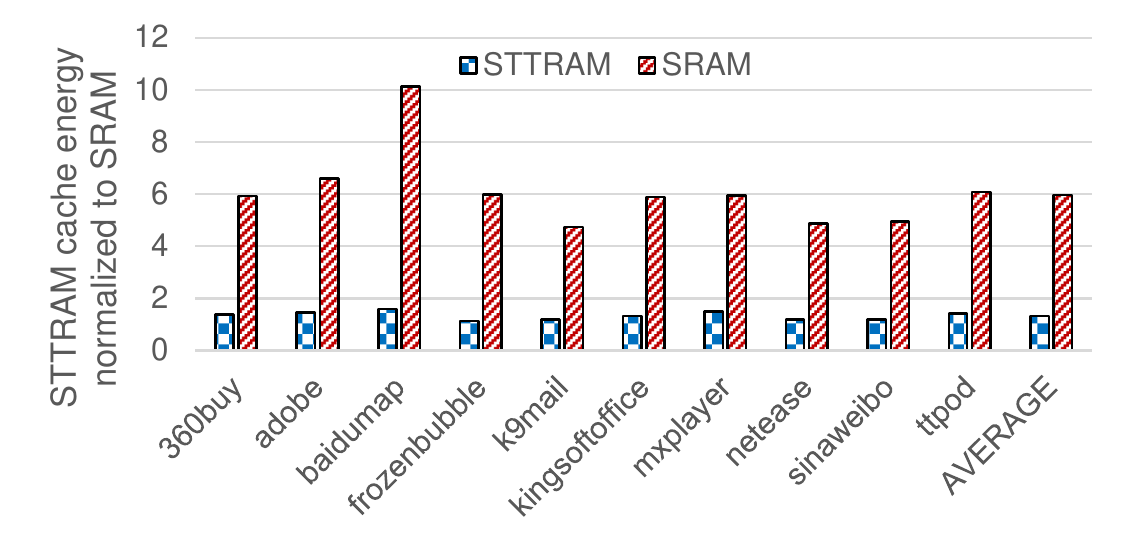}
		\vspace{-7pt}
		\caption{Total cache energy of quad-core system with L2 normalized to system without L2.}
		\label{fig:l2-compare}
        \vspace{-10pt}
\end{figure}

For most applications, the presence of the L2 cache increased the energy compared to the system without the L2 cache. However, a notable difference was that the STTRAM-based systems suffered substantially less energy overhead from the presence of the L2 cache than the SRAM-based systems. Figure \ref{fig:l2-compare} illustrates this observation. The figure depicts the total cache energy consumption of the quad-core system with L2 cache normalized to the quad-core system without the L2 cache, for both STTRAM and SRAM. For brevity, the STTRAM results represent the averages of the energy of the 1$ms$, 10$ms$, and 100$ms$ retention times, since the trends were similar for the different retention times. On average across all the benchmarks, for the STTRAM-based system, introducing the L2 cache increased the energy by 32.4\%. For the SRAM-based system, on the other hand, introducing the L2 cache increased the energy by 6.0x. Even though the L2 cache did not introduce as much energy in the STTRAM-based system as it did in the SRAM-based system, system designers must carefully consider the tradeoffs of including an L2 cache, especially for energy-constrained systems.

\section{Asymmetric STTRAM Cache Design}
The results presented so far have shown that STTRAM can provide substantial energy benefits over SRAM for mobile applications. We have also shown that while STTRAM performed well for all the mobile applications, the benefits varied for different applications. This observation is in line with prior research that has shown that different applications may have different cache requirements. Thus, to further improve the energy efficiency of STTRAM caches for mobile applications, we propose \textit{asymmetric retention time cache} as a simple design approach for mobile device processors. 

\subsection{Proposed Retention Time Asymmetry}
\begin{figure}[t]
		\vspace{-10pt}
		\centering
		\includegraphics[width=0.7\linewidth]{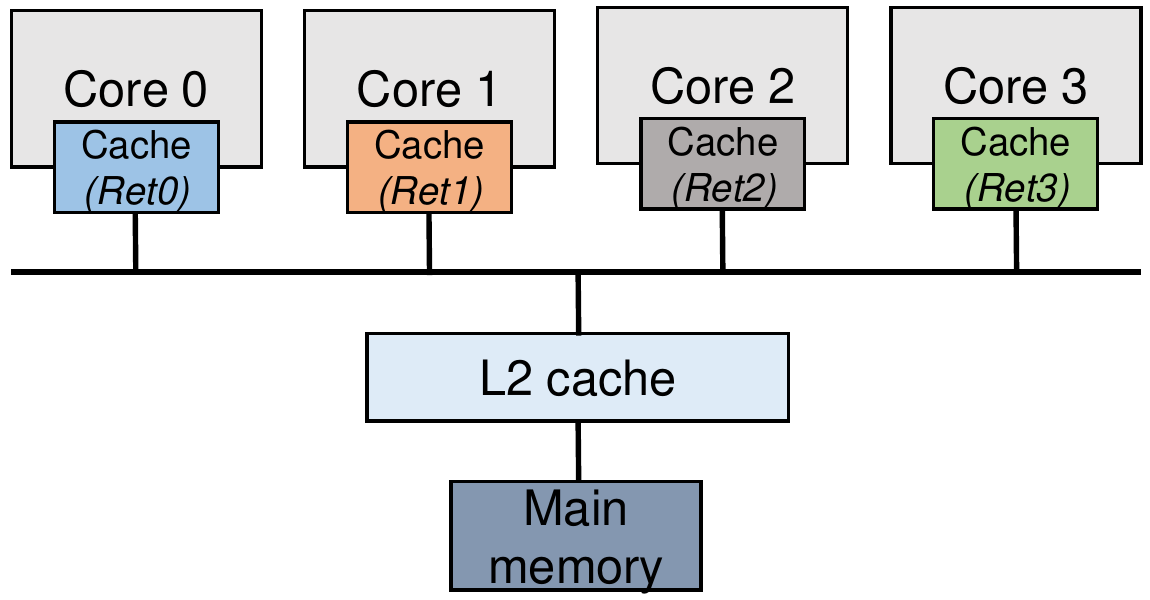}
		\caption{Illustration of proposed asymmetric retention caches. The L1 caches feature different retention times to satisfy a variety of execution requirements.}
		\label{fig:asymmetric}
        \vspace{-10pt}
\end{figure}

Figure \ref{fig:asymmetric} illustrates the proposed asymmetric retention STTRAM cache design. The multicore processor is designed using different retention times in each core such that threads requiring similar retention times can be scheduled to their best core during runtime by the operating system. In the private L1 instruction and data caches, different cores feature different retention times, which are carefully chosen via design-time exploration, to satisfy a variety of applications' execution requirements. Additional asymmetry can also be implemented in lower cache levels. For example, shared caches (e.g., L2, L3) can be designed with different retention times in different banks in a multi-banked design, and cache blocks are opportunistically written to the bank that most closely matches the blocks' lifetimes \cite{kuan19}. In the following subsection, for brevity, we limit our evaluations to the benefits of asymmetric L1 STTRAM cache design.

\subsection{Asymmetric Retention L1 STTRAM Cache}
In the asymmetric L1 STTRAM cache, during runtime, threads are scheduled to the core that most closely matches the threads' execution requirements as determined during a profiling period. Unlike the scenario described in Section \ref{sec:retSpecialization} where different applications are run using different retention times on a single chip, in the asymmetric setup, different threads of the same application---or different applications in a multi-programmed workload---are run on the cores with retention times that most closely satisfy the threads' needs.   

\begin{figure}[b]
		\vspace{-25pt}
		\centering
		\includegraphics[width=0.7\linewidth]{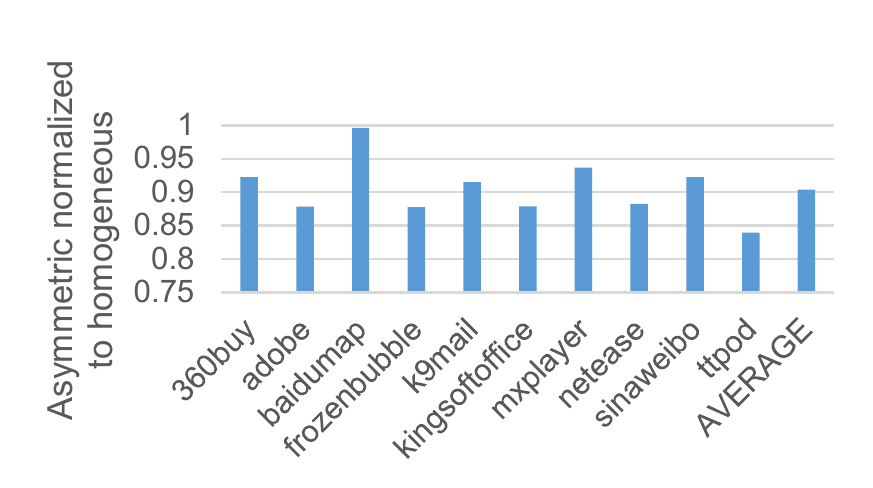}
		\vspace{-7pt}
		\caption{Energy consumption of asymmetric retention times normalized to the best homogeneous retention time.}
		\label{fig:asymmetric-E}
\end{figure}

Overall, asymmetric retention L1 STTRAM cache achieved energy savings compared to a base homogeneous L1 STTRAM cache without degrading the execution time. Figure \ref{fig:asymmetric-E} compares the asymmetric retention design to a base homogeneous retention time (baseline of one). The base homogeneous retention time is chosen as the single best retention time that achieved the lowest energy on all the cores for each benchmark. For a stringent comparison, we used application-specific homogeneous retention times; thus, the homogeneous configuration changed for different benchmarks. On average, the asymmetric design reduced the energy by 9.6\% compared to the base homogeneous retention time, with energy savings as high as 16.0\% for \textit{ttpod}. In the worst case, the asymmetric design performed similarly to the homogeneous design for \textit{baidumap}. These results suggest that that the asymmetric design is a viable approach to further improve the energy efficiency of STTRAM caches for mobile applications. 

\subsection{Profiling for Retention Time} \label{sec:profiling}
To fully leverage the benefits of an asymmetric L1 STTRAM cache design, one important challenge is how to determine the best retention time during runtime. Solving this challenge is outside the scope of this paper; however, we suggest a few potential solutions, some of which have been explored by related work. 

The first potential technique for determining the best retention time is a sampling technique as described in prior work \cite{kuan18}. The executing application is run on each core for a brief profiling interval (e.g., 10M instructions) after which the energy consumption is calculated. The rest of the application is then run on the core whose cache consumed the least energy. Sampling can be used without introducing substantial overheads if only a few retention times (e.g., four) must be sampled. However, in larger-scale systems, other techniques must be explored. 

Unlike traditional cache configurations (cache size, line size, and associativity), the best retention time cannot be directly predicted using statistics obtained from performance counters (e.g., cache miss rates). Thus, an alternative solution is to use design-time machine learning algorithms to model the correlation of retention time requirements to easily obtainable statistics. These models can then be used during runtime to directly predict the best retention time, given the execution statistics obtained during a profiling interval. We plan to evaluate these solutions for mobile applications in future work.


\section{Concluding Remarks}
Spin-Transfer Torque RAMs (STTRAMs) have the potential to replace SRAMs in implementing on-chip caches in emerging mobile devices. Reduced retention STTRAMs, in addition to area benefits, offer further energy savings potential, if the retention times are carefully chosen to satisfy the cache block lifetimes of executing applications. In this paper, we explore and evaluate the benefits of STTRAM caches for mobile applications. We characterize mobile applications from the perspective of reduced retention STTRAM caches, and show that STTRAMs provide much energy benefits for mobile applications without introducing substantial latency overheads. We also explored the benefits of an asymmetric retention time design to provide further energy savings compared to a homogeneous retention time design. The analysis herein is performed without any orthogonal techniques for making STTRAMs more efficient (e.g., limiting the number of writes, hybrid caches), thereby minimizing runtime implementation overheads. However, our future work includes a detailed study of the synergy of prior techniques for energy efficient cache hierarchies for mobile applications and evaluating different STTRAM cell models for emerging mobile devices.

	\balance
	\bibliographystyle{IEEEtran}
	\bibliography{refs}
	
\end{document}